# Magnetization reversal and current hysteresis due to spin injection in magnetic junction.

Roger J. Elliott,[a] Ernest M. Epshtein,[b] Yuri V. Gulyaev,[b] Peter E. Zilberman[b*]

[a] *University of Oxford, Department of Physics, Theoretical Physics, 1 Keble Road, Oxford OX1 3NP, United Kingdom*
[b] *Institute of Radio Engineering and Electronics of the Russian Academy of Sciences, Fryazino Branch, Vvedenskii sq. 1, Fryazino, Moscow District, 141190, Russia.*

**Abstract**

Magnetic junction is considered which consists of two ferromagnetic metal layers, a thin nonmagnetic spacer in between, and nonmagnetic lead. Theory is developed of a magnetization reversal due to spin injection in the junction. Spin-polarized current is perpendicular to the interfaces. One of the ferromagnetic layers has pinned spins and the other has free spins. The current breaks spin equilibrium in the free spin layer due to spin injection or extraction. The nonequilibrium spins interact with the lattice magnetic moment via the effective s-d exchange field, which is current dependent. Above a certain current density threshold, the interaction leads to a magnetization reversal. Two threshold currents are found, which are reached as the current increases or decreases, respectively, so that a current hysteresis takes place. The theoretical results are in accordance with the experiments on magnetization reversal by current in three-layer junctions Co/Cu/Co prepared in a pillar form.

## 1. Introduction

Magnetization reversal phenomenon arising due to spin-polarized current in magnetic junctions has been discussed for the first time by Slonczewski and Berger in the papers [1, 2]. This phenomenon has been observed in a number of experimental works (see [3 - 5]). According to [1, 2] a torque arises acting on magnetic lattice when mobile electrons in current intersect a boundary between two metallic ferromagnetic layers of the junction. It occurs because magnetization vectors of the ferromagnets form in general some angle between each other, which is different from $0$ or $\pi$. In the case electrons are forced to adapt themselves for new quantization axis. This principally quantum mechanical process acts very near the interface in a range of a few electron wavelengths, typically ~ 1 - 2 nm.

Under the discussion of the mechanism some other significant aspect turned out in a shadow. As it was firstly pointed out in the papers [6 - 8] an interesting phenomenon arises after the current electrons penetrate apart from the interface into the bulk of the ferromagnetic layer. This question was stated in the papers [9 - 11] also.

We continue here to develop the theoretical approach proposed firstly in the paper [10] and show a novel spin-injection mechanism may act deep in the bulk of the ferromagnetic layer in a range of a spin-diffusion length $l$. The latter one may be large enough, namely $l \sim$ 10 - 100 nm for ferromagnetic metals. We believe this bulk mechanism is a complementary one to the mechanism proposed and discussed in the papers [1, 2, 12, and 13]. In our opinion, it is a very significant task now to compare all the proposed mechanisms with the experimental data in details. We present here additional results of the spin injection mechanism theory and show the results apparently may be in accordance with some experimental data reported for three-layer junctions Co/Cu/Co prepared in the pillar form [3 - 5].

We consider a conventional model of a magnetic junction with two ferromagnetic layers one of which (layer 1) is pinned and the other (layer 2) is free. Very thin spacer layer exists between the layers 1 and 2. The layer 2 has in our model finite thickness and contacts at the second end with some nonmagnetic conductor (layer 3). The spin-polarized current may inject (or extract) nonequilibrium spins into layer 2. These spins interact with the lattice magnetization due to effective s-d exchange field that depends on the current. At a sufficiently large current density, the effective field leads to the magnetization reversal of the layer 2.

## 2. Nonequilibrium spin polarization

We consider further only stable stationary state of the junction when a steady current is flowing perpendicular to the interfaces. Our aim is to find the threshold currents giving frontiers of the stable behavior. Let us mark out a region of the layer 2 adjoining to 1, 2 layers interface. Inside this region mobile electrons (or current carriers) flowing from layer 1 to layer 2 adapt themselves for new quantization axis that is for the direction of the lattice magnetization vector $\mathbf{M}^{(2)}$. Thickness of the region may be estimated as $\lambda_J \leq 1-2$ nm [1, 2, 9, 11 – 13]. It means thus the thickness is much smaller than the spin diffusion length $l \sim 10-100$ nm for current carriers. The region was introduced firstly by Slonczewski and Berger [1, 2] and will be referred to further as "SB layer". As it was shown in the works cited a torque arises inside this thin region due to transversal components of the mobile electron spins ("SB torque"). This torque is current dependent and for large enough current densities $j > j_{th}^t$ may overcome dissipative processes in the whole layer 2 and distort the stability (lead to switching).

Let us return now to smaller currents and stable stationary state of the junction. After leaving the thin region in the vicinity of the interface, electrons become longitudinally polarized and stationary distributed over the energy spin subbands of the layer 2. The distribution becomes stationary but not equilibrium because of the current. This phenomenon is analogous to the well known one in semiconductor physics and will be called further as "spin injection" due to current flowing from layer 1 to layer 2. The injected spins occupy the whole space of the layer 2 up to the distances of the diffusion length $l$ and for more distances spin equilibrium restores. We will try show further in the paper the longitudinal nonequilibrium injected spins do interact strongly with the lattice magnetization $\mathbf{M}_2$ due to s-d exchange. Corresponding "effective field" is, of course, current dependent. This effective field disturbs the magnetic stability and leads to magnetization reversal of the layer 2 for large enough current densities $j > j_{th}^l$ and for some appropriate additional conditions. As it will be shown the threshold current $j_{th}^l$ is not sensitive to dissipation processes. This property is in contrast with the well known proportionality of the threshold $j_{th}^t$ to dissipation parameter. It allows us to neglect completely any dissipation in the magnetic subsystem when calculating $j_{th}^l$. The problem arises: what of two mentioned mechanisms of stability losing acts in reality. It depends on the threshold $j_{th}^l$, whether it will be higher or lower than the $j_{th}^t$.

To calculate $j_{th}^l$ we start now to describe spin injection in the layer 2. It may be done by means of introducing the function

$$P^{(2)}(x) = \frac{n_\uparrow^{(2)}(x) - n_\downarrow^{(2)}(x)}{n}, \qquad (1)$$

where $x$ is the coordinate along the current, originated in the 1, 2 layer interface, $n_{\uparrow,\downarrow}^{(2)}(x)$ are the densities of carriers with opposite spins, and $n^{(2)} = n_\uparrow^{(2)}(x) + n_\downarrow^{(2)}(x)$ is the total carrier density independent on the coordinate $x$ because of the metal charge neutrality condition. Function $P^{(2)}(x)$ is the spatial distribution of the spin polarization degree. This function is to obey in the layer 2 the steady-state diffusion equation

$$\frac{d^2 P^{(2)}(x)}{dx^2} - \frac{P^{(2)}(x) - \overline{P}^{(2)}}{l^{(2)2}} = 0, \qquad (2)$$

where $\overline{P}^{(2)}$ denotes the equilibrium value of $P^{(2)}(x)$, $l^{(2)}$ is the spin diffusion length in the layer 2. The drift term is omitted here because it is negligibly small in metals for current density range used (for more details see [10]). Analogous equation may be written for the layer 3 also.

Equation (2) is valid for description of electrons having only longitudinal spin polarization. Such electrons appear far enough from the interface, namely, outside the SB layer with $\lambda_J$ in thickness. However, SB layer itself plays an important role also: it determines the true choice of the solution of the equation (2). This role may be ensured by means of appropriate boundary conditions.

To derive the conditions let us start from the following considerations. As it is well known (see e.g. [14]), any spin parallel to the quantization axis $\mathbf{z}^{(1)}$ (in our case $\mathbf{z}^{(1)} \parallel \mathbf{M}^{(1)}$, $|\mathbf{z}^{(1)}| = 1$ and $\mathbf{M}^{(1)}$ is the magnetization vector of the layer 1) has the probability $\cos^2 \varphi/2$ be parallel to the other axis $\mathbf{z}^{(2)} \parallel \mathbf{M}^{(2)}$ and the probability $\sin^2 \varphi/2$ be antiparallel to this axis. Here and further we designate by $\varphi$ an angle between axes $\mathbf{z}^{(1)}$ and $\mathbf{z}^{(2)}$. For the spin, which is antiparallel to $\mathbf{z}^{(1)}$, the expressions for probabilities indicated should interchange each other, that is $\cos^2 \varphi/2 \leftrightarrow \sin^2 \varphi/2$.

Just as any electron intersects the boundary between the layers 1 and 2 it occurs in a nonstationary quantum state and the probability to be present in any of two spin energy subbands of the layer 2 varies with time. In classical terms it means that a transversal component of the total mobile electron spin arises, performs a precession and creates the SB torque. But after leaving the SB layer this nonstationary situation ends and electrons become distributed over the energy spin subbands in accordance with the probabilities mentioned above [15].

Let us consider now an explicit expression for the partial current densities in spin subbands of the $i$th layer ($i = 1, 2, 3$)

$$j_{\uparrow,\downarrow}^{(i)}(x) = e\mu_{\uparrow,\downarrow}^{(i)} n_{\uparrow,\downarrow}^{(i)}(x) E^{(i)}(x) - eD_{\uparrow,\downarrow}^{(i)} \frac{dn_{\uparrow,\downarrow}^{(i)}(x)}{dx}, \quad (3)$$

where $\mu_{\uparrow,\downarrow}^{(i)}$, $D_{\uparrow,\downarrow}^{(i)}$, $E^{(i)}(x)$ are the electron mobility, diffusion coefficient and electric field strength, respectively. For the current flowing in the direction $1 \to 2$ we may write the following conditions near the 1, 2 interface

$$j_{\uparrow,\downarrow}^{(2)} = j_{\uparrow,\downarrow}^{(1)} \cdot \cos^2 \varphi/2 + j_{\downarrow,\uparrow}^{(1)} \cdot \sin^2 \varphi/2. \quad (4)$$

The conditions (4) are in accordance with our previous considerations about the probabilities of electron distribution over the spin subbands. Indeed, any partial current in the layer 2 is a sum of the contributions from the two partial currents of the layer 1, taken with appropriate probabilities.

Let us introduce now spin current by the expression

$$J_s^{(i)}(x) \equiv \frac{\hbar}{2e} \cdot \left[ j_\uparrow^{(i)}(x) - j_\downarrow^{(i)}(x) \right]. \quad (5)$$

If we substitute the conditions (4) into formula (5), we obtain near the 1, 2 interface

$$J_s^{(2)} = J_s^{(1)} \cdot \cos \varphi. \quad (6)$$

The condition of spin current continuity in the form (6) has been derived by different way in [10]. In our case, however, it is more convenient to rewrite (6) in terms of the function $P^{(2)}(x)$. We can express densities $n_{\uparrow,\downarrow}^{(2)}$ via this function using (1) and neutrality condition. It gives

$$n_{\uparrow,\downarrow}^{(2)}(x) = \frac{n}{2} \cdot \left[ 1 \pm P^{(2)}(x) \right]. \quad (7)$$

We can use further the expression (3) and the condition of the total current $j = j_\uparrow^{(2)} + j_\downarrow^{(2)}$ continuity to exclude the field strength $E^{(2)}(x)$ and express it via the densities and after that via the $P^{(2)}(x)$ by means of (7). We get finally after direct calculations

$$\left. \frac{d\Delta P}{dx} \right|_{x=+0} = -\frac{Q^{(1)} \cos \varphi - Q^{(2)}}{l^{(2)}} \frac{j}{j_D^{(2)}}. \quad (8)$$

where the nonequilibrium part of the function $P^{(i)}(x)$ is $\Delta P(x) \equiv P^{(i)}(x) - \overline{P}^{(i)}$ and some critical current is introduced $j_D^{(2)} \equiv en^{(2)} l^{(2)} / \tau^{(2)}$, $\tau^{(2)}$ is the spin relaxation time in the layer 2, $l^{(2)} = \sqrt{\widetilde{D}^{(2)} \cdot \tau^{(2)}}$. Effective diffusion coefficient is

$$\widetilde{D}^{(2)} = \frac{D_\uparrow^{(2)}\mu_\downarrow^{(2)}\overline{n}_\downarrow^{(2)} + D_\downarrow^{(2)}\mu_\uparrow^{(2)}\overline{n}_\uparrow^{(2)}}{\mu_\downarrow^{(2)}\overline{n}_\downarrow^{(2)} + \mu_\uparrow^{(2)}\overline{n}_\uparrow^{(2)}}. \quad (9)$$

Effective polarization parameters in (8) are defined by:

$$Q^{(i)} = \frac{\mu_\uparrow^{(i)} n_\uparrow^{(i)} - \mu_\downarrow^{(i)} n_\downarrow^{(i)}}{\mu_\uparrow^{(i)} n_\uparrow^{(i)} + \mu_\downarrow^{(i)} n_\downarrow^{(i)}}. \quad (10)$$

Note that typically $j_D^{(2)} \sim 10^{10}\,\text{A/cm}^2$. We will see further that $j \sim j_{th}^l \leq 10^7 - 10^8\,\text{A/cm}^2$. Therefore the ratio $j/j_D^{(2)}$ in (8) is a very small parameter.

Boundary condition for gradients of $\Delta P$ at the interface $x = L$ between the ferromagnetic layer 2 and a nonmagnetic layer 3 may be derived similarly and takes the following form:

$$j_D^{(2)} l^{(2)} \left.\frac{d\Delta P}{dx}\right|_{x=L-0} - j_D^{(3)} l^{(3)} \left.\frac{d\Delta P}{dx}\right|_{x=L+0} = Q_2 j, \quad (11)$$

where $L$ is the layer 2 thickness.

The second boundary condition for $\Delta P$ is needed at the interface $x = L$. The condition may be derived from the chemical potential continuity [16, 17]. It takes the form:

$$N^{(2)} \Delta P\big|_{x=L-0} = N^{(3)} \Delta P\big|_{x=L+0}, \quad (12)$$

where

$$N^{(i)} = \zeta_0^{(i)} \left[ \left(1 + \overline{P}^{(i)}\right)^{-1/3} + \left(1 - \overline{P}^{(i)}\right)^{-1/3} \right], \quad (13)$$

and $\zeta_0^{(i)} = \frac{(3\pi^2 n^{(i)})^{2/3} \hbar^2}{2m}$, $m$ is an electron effective mass.

Solving Eq. (2) and analogous equation in the layer 3 with boundary conditions (8) - (11), we obtain the following nonequilibrium spin polarization in the layer 2:

$$\Delta P(x) = \frac{j}{j_D^{(2)}} (\sinh\lambda + \nu\cosh\lambda)^{-1}$$
$$\times \left\{ Q^{(2)} \cosh\lambda + \left(Q^{(1)}\cos\varphi - Q^{(2)}\right) \right.$$
$$\left. \times \left[\cosh(\lambda - \xi) + \nu\sinh(\lambda - \xi)\right] \right\}, \quad (14)$$

where $\lambda = L/l^{(2)}$, $\xi = x/l^{(2)}$,
$\nu = \left(j_D^{(3)}/j_D^{(2)}\right)\left(N^{(2)}/N^{(3)}\right)$. Parameter $\nu$ describes influence of the nonmagnetic layer 3.

## 3. Magnetic energy

Let us suppose the easy axis of the layer 2 lies in the layer plane and makes an angle $\beta$ with the quantization axis $\mathbf{z}^{(1)}$ of the layer 1. Then, the anisotropy energy per unit area of the layer 2 is

$$U_A = -KL\cos^2(\varphi - \beta), \quad (15)$$

where $K$ is the anisotropy constant.

If an external magnetic field $\mathbf{H}$ is applied in the junction plane at an angle $\gamma$ to the axis $\mathbf{z}^{(1)}$, then the Zeeman energy in the layer 2 is

$$U_H = -\widetilde{M}^{(2)} HL\cos(\varphi - \gamma), \quad (16)$$

where $\widetilde{M}^{(2)} = M^{(2)} + \mu_B n^{(2)} \overline{P}^{(2)}$ is the equilibrium magnetization of the layer 2 produced by localized magnetic moments and free carrier spins and $\mu_B$ is the Bohr magneton.

The injected spins contribute to the magnetic energy of the sample via the *s-d* exchange interaction with the lattice magnetic moment. This energy per unit area is

$$\Delta U_{s-d} = -\alpha\mu_B n^{(2)} M^{(2)} \int_0^L \Delta P(x) dx, \quad (17)$$

where $\alpha$ is the dimensionless *s-d* exchange interaction constant, its typical value being $\sim 10^4 - 10^6 \gg 1$.

Substituting (14) into (17) we obtain
$$\Delta U_{s-d} = -\alpha\mu_B n^{(2)} M^{(2)} l^{(2)}$$
$$\times \frac{j}{j_D^{(2)}} (\sinh\lambda + \nu\cosh\lambda)^{-1}$$
$$\times \left[ Q^{(1)}\cos\varphi \sinh\lambda + \nu\left(Q^{(1)}\cos\varphi - Q^{(2)}\right)(\cosh\lambda - 1) \right]. \quad (18)$$

The current density through the junction depends on the angle $\varphi$. This dependence is determined by the transformation rules for spin wave functions under rotation of the quantization axis [18] and has the form
$$j = \tfrac{1}{2}(j_p + j_a) + \tfrac{1}{2}(j_p - j_a)\cos\varphi, \quad (19)$$

where $j_p$ and $j_a$ are the current densities for the parallel ($\varphi = 0$) and antiparallel ($\varphi = \pi$) relative orientation of the magnetizations, respectively. An external magnetic field can change the angle $\varphi$, which leads to the effect of giant magnetoresistance; its measure is the ratio $\rho \equiv (j_p - j_a)/j_p$.

In view of Eq. (19), the total angular dependence of the s-d exchange energy takes the form

$$\Delta U_{s-d} = A - B\cos\varphi - C\cos^2\varphi, \qquad (20)$$

where

$$A = G(j_p + j_a)Q^{(2)}\nu(\cosh\lambda - 1), \qquad (21)$$

$$B = G\{(j_p + j_a)Q^{(1)}[\sinh\lambda + \nu(\cosh\lambda - 1)]$$
$$- (j_p - j_a)Q^{(2)}\nu(\cosh\lambda - 1)\}, \qquad (22)$$

$$C = G(j_p - j_a)Q^{(1)}$$
$$\times [\sinh\lambda + \nu(\cosh\lambda - 1)], \qquad (23)$$

$$G = \frac{\alpha\mu_B n^{(2)} M^{(2)} l^{(2)}}{2 j_D^{(2)}(\sinh + \nu\cosh\lambda)}. \qquad (24)$$

It is seen from a comparison of (20) with (15) and (16) that the *s-d* exchange energy is equivalent to the appearance of an additional magnetic field $H' = B/\widetilde{M}^{(2)}L$, which is parallel to the $\mathbf{z}^{(1)}$ axis, and that of an additional anisotropy with an anisotropy constant $K' = C/L$ and an anisotropy axis parallel to the $\mathbf{z}^{(1)}$ axis. It is known that the changes in the field or in the anisotropy constants can lead to reorientation phase transitions in magnetic films. We will show further the following: as the spin-polarized current leads to the additional field $H'$ and the additional constant $K'$, it may also lead under certain conditions to the magnetization reversal of the layer 2.

## 4. Magnetization reversal by current

To calculate further it is convenient for us to use energy minimum principle. Let us consider the magnetic subsystem with the Hamiltonian

$$U = U_{s-d} + U_A + U_H, \qquad (25)$$

considered as a function of the angle $\varphi$. As it is justified in Appendix our subsystem is in a partial equilibrium state in spite of the presence of a current. It is because the current do not produce any irreversible processes in the subsystem. The equilibrium value of the angle $\varphi$ may be therefore determined from the minimum conditions (A6) and (A10). Let us consider the simplest and, apparently, the most important case where $\beta = 0$ and $\gamma = \pi$. The calculations become identical, in principle, to those used in the conventional theory of magnetization reversal by external magnetic field (see, for instance, [19]). Therefore, the magnetic state is determined by a current-dependent parameter $\eta \equiv \dfrac{\widetilde{M}^{(2)}HL - B}{2(KL + C)}$. At $|\eta| < 1$, two equilibrium angular states exist, $\varphi = 0$ and $\varphi = \pi$, one of them being stable while the other is metastable; at $\eta > 1$, there is only one equilibrium angle, $\varphi = \pi$; and at $\eta < -1$, there is only one angle, $\varphi = 0$. The magnetization switching from $\varphi = \pi$ to $\varphi = 0$ or vice versa occurs when the metastable state disappears. Existence of the metastable states leads to hysteresis behavior of the magnetization switching.

Let us take initially current $j = 0$ and a sufficiently high magnetic field $|H| > 2K/\widetilde{M}^{(2)}$ directed opposite to the layer 1 magnetization. Such a situation corresponds to $\eta > 1$, so that $\varphi = \pi$. If the current increases to the positive direction, i.e. $j/j_D^{(2)} > 0$, the initial magnetization state is retained until $\eta = -1$. Then the switching to $\varphi = 0$ occurs. The threshold current density for such a process, $j_{a\to p}$, can be found from the condition $\eta = -1$. We get from the condition using formulae (22) and (23) the following expression

$$j^t{}_{a\to p} = j_0(h + h_A)(1-\rho)\lambda(\sinh\lambda + \nu\cosh\lambda)$$

$$\times \{Q^{(1)}[\sinh\lambda + \nu(\cosh\lambda - 1)](2 - 3\rho)$$

$$- Q^{(2)}\nu(\cosh\lambda - 1)\rho\}^{-1}, \qquad (26)$$

where

$$h = \frac{H}{4\pi\widetilde{M}^{(2)}}, \quad h_A = \frac{K}{2\pi\widetilde{M}^{(2)2}},$$

$$j_0 = j_D^{(2)} \frac{8\pi \widetilde{M}^{(2)2}}{\alpha \mu_B n^{(2)} M^{(2)}}.$$

If the current changes to the opposite direction, the switching from $\varphi = 0$ to $\varphi = \pi$ occurs at $\eta = 1$. The corresponding threshold current density is

$$j^l_{p \to a} = j_0 (h - h_A) \lambda (\sinh \lambda + \nu \cosh \lambda)$$

$$\times \{Q^{(1)}[\sinh \lambda + \nu(\cosh \lambda - 1)](2 + \rho)$$

$$- Q^{(2)} \nu (\cosh \lambda - 1) \rho\}^{-1}. \qquad (27)$$

Remember that in accordance with our previous denomination (see section **2**) the upper index "*l*" in the formulae (26) and (27) should stress the thresholds are determined by the longitudinal spins of mobile electrons. It follows from Eqs. (26) and (27) that the threshold currents rise linearly with $L$ at $L \gg l^{(2)}$. Therefore, $L \leq l^{(2)}$ condition corresponds to minimal switching threshold.

Comparing formulae (26) and (27), we see that the threshold currents do not coincide. What threshold is working depends on the direction of the current changes. In other words, a current hysteresis takes place. Similar situation was observed experimentally [3 - 5] upon investigations of pillar magnetic junctions Co/Cu/Co. In this connection, we perform further a more detailed comparison of our calculations with the results of these works.

## 5. Comparison with the experimental data

It is convenient to represent the calculated dependence of the junction differential resistance $dV/dj$ on the current density $j$, where $V$ is the voltage of the current source. Naturally, the result will depend on the conduction mechanism in the junction (ohmic conduction, ballistic transport, heating effects, etc.). For simplicity, we assume the ohmic conduction. Then $dV/dj = R + r(\varphi)$, where $R$ is the internal resistance of the source, and $r(\varphi)$ is the junction resistance depending on the angle $\varphi$. These dependencies for $\varphi = \pi$ and $\varphi = 0$ are shown in the fig.1. Arrows indicate directions of the changes in resistance depending on the current. The resistance jumps related to magnetization reversal are seen. Qualitatively, such dependence corresponds completely to experimental data [3, 4].

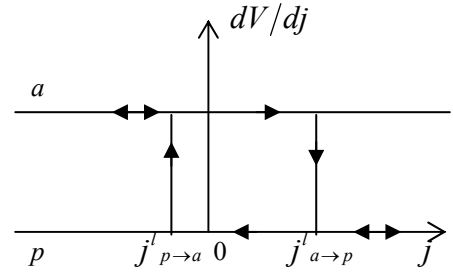

Fig. 1

For quantitative estimations, we take the following parameters for Co films: $M \sim 0.1$ T, $K \sim 0.4$ J/cm$^3$, $\alpha \sim 2 \times 10^5$, $n \sim 10^{22}$ cm$^{-3}$, $\tau \sim 10^{-13}$ s, $l^{(2)} \sim L \sim 10^{-6}$ cm, $Q^{(1)} = 0.35$, $Q^{(2)} = 0.2$. The magnetoresistance ratio can be written as $\rho = (r(\pi) - r(0))/(R + r(\pi))$. In experiments [3, 4] the relation $R \gg r(\theta) \geq r(0)$ was fulfilled, that is the magnetoresistance ratio was very small. Therefore, upon the estimation of threshold currents by formulae (26) and (27) we assume $\rho = 0$. Then we obtain (at $h = 0$) $j^l_{a \to p} = -j^l_{p \to a} \sim 10^7$ A/cm$^2$, which approximately corresponds to the experimental estimations of the threshold currents. We emphasize here that, as in experiments, two threshold currents have different signs and coincide in absolute magnitude at $h = 0$. In complete accordance with experiments, the thresholds are shifted in the positive direction under the applied magnetic field $h$, and the symmetry in the location of the two thresholds with respect to the point $j = 0$ is broken down (see fig. 1). The theory predicts that one of the thresholds vanishes, namely, $j^l_{p \to a} = 0$, at $h = h_A$. At above-mentioned values of the parameters, this corresponds to field $H = 0.8$ T.

## 6. Discussion

We would stress now some distinguishing features of the "effective field" or "spin- injection" mechanism that may help to identify it in the experiments:

1. The thresholds $j^l_{th}$ (that is $j^l_{p \to a}$ or $j^l_{a \to p}$) do not sensitive to dissipation parameter $\alpha_d$. According to formulae (26) and (27) these threshold currents do not depend on $\alpha_d$ but critically depend on anisotropy field $h_A$.

2. Let us consider now thickness dependence of the thresholds. We suppose the minimal thresholds are satisfied the relation

$$\min(j^t_{th}) < \min(j^l_{th}). \qquad (28)$$

Experiment [20] and theories (see e.g. our formulae (26) and (27)) give the thresholds rise linearly with increasing of the layer 2 thickness $L$. Scheme in the fig. 2 illustrates the dependencies qualitatively.

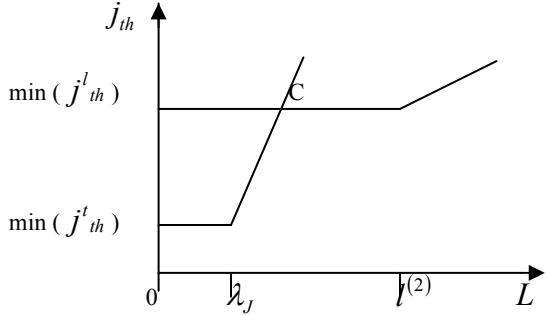

Fig. 2

Note that slope of these linear dependences is to be proportional to the ratio of the total layer 2 volume and of the volume of its active part. For $j^t_{th}$ calculation the active part is just one where the torque acts and has therefore $\lambda_J$ in thickness. For $j^l_{th}$ calculation the active part corresponds to the region of effective field action that is to the diffusion length $l^{(2)}$. As typically the relation $l^{(2)} \gg \lambda_J$ is valid, we may expect the slope of $j^t_{th}$ be greater than the slope of $j^l_{th}$. Just the situation is shown in the fig. 2. If we take large enough thickness $L$ an intersection point $C$ may appear. On the right hand side of the point the threshold $j^l_{th}$ becomes lower and may therefore determine the instability onset.

Some recent experiments [21, 22] considered only structures having very small $L \approx 2.5$ nm. Those are the most favorable structures to observe SB torque and authors conclude they do observe it. Experiment [20] shows the linear dependence of the threshold exists on the thickness $L$. Just linear dependence follows directly from our theory formulae (26) and (27). More information may be found experimentally from slope investigations. In particular, the slope does not depend at all on the dissipation for this longitudinal spin injection mechanism.

We would note in the conclusion that there are no theory grounds to rule out the longitudinal spin injection mechanism. Our numerical estimations of magnitude and sign of the threshold current $j^l_{th}$ as well as its dependence on the field $H$ and thickness $L$ show close correspondence of the theory with the experimental data. Therefore the longitudinal spin injection should be taken into account in the further investigations as an additional possibility to switch magnetic junctions by spin polarized current.

### Acknowledgements

The work was supported by International Scientific & Technology Center (grant ISTC # 1522) and by Russian Foundation on Basic Research (grant RFBR # 03-02-17540).

### APPENDIX

**Equivalence of dynamic equations and energy minimization principle in the problem of a steady current driven switching in magnetic junctions.**

Our consideration here refers to the layer 2 only. Therefore we omit further any indications on the layer number. General dynamics of the lattice magnetization may be described by the well known Landau-Lifshitz-Gilbert equation [23]

$$\frac{d\mathbf{M}}{dt} = -\gamma \mathbf{M} \times \mathbf{H}_{\text{eff}} + \frac{\alpha_d}{M} \cdot \mathbf{M} \times \frac{d\mathbf{M}}{dt}, \qquad (A1)$$

where $\gamma > 0$ is a gyromagnetic ratio, $\alpha_d > 0$ is a damping parameter, $t$ is a time and the total effective field (according to its definition) is equal to

$$\mathbf{H}_{eff} = -\frac{\partial U_{tot}}{\partial \mathbf{M}}, \qquad (A2)$$

where the $U_{tot} = U + 2\pi M_x^2$ is the total magnetic energy including (26) and demagnetization energy (the last term). We want stress here that the total magnetic energy is dependent on the current density $j$ (see (18)). Therefore the whole system is not in equilibrium state. But our system is in stationary state and we may write therefore $d\mathbf{M}/dt = 0$. In such a case the equation (A1) reduces to

$$\mathbf{M} \times \frac{\partial U}{\partial \mathbf{M}} = 0. \qquad (A3)$$

Remember that in stationary state vector $\mathbf{M}$ lies in the *yz* plane, which is parallel to interface, and $\mathbf{M} \parallel \mathbf{z}$, axis *x* being parallel to the current. We lay therefore $M_x = 0$ and replace $U_{tot} \to U$ in (A3). To describe the stationary state we introduce the following polar components of the magnetization

$$M_y = M \cdot \sin\varphi, \quad M_z = M \cdot \cos\varphi. \qquad (A4)$$

Vector product (A3) has *x*-component only. Therefore it may be written as

$$[M_y \cdot \left(\frac{\partial M_z}{\partial \varphi}\right)^{-1} - M_z \cdot \left(\frac{\partial M_y}{\partial \varphi}\right)^{-1}] \cdot \frac{\partial U}{\partial \varphi} = 0 \qquad (A5)$$

We use (A4) to calculate derivatives in (A5). Then the Eq. (A5) may be written as

$$\frac{\partial U}{\partial \varphi_0} = 0, \qquad (A6)$$

where we denominate $\varphi = \varphi_0$ all roots of (A6). The latter equation shows any stationary solution may be found whether from the equation (A3) or from the energy condition (A6). These two ways are equivalent completely.

The next step is to separate stable and unstable solutions of (A6). If angle $\varphi$ differs slightly from $\varphi_0$ it becomes time dependent and should be described by the general Eq. (A1). It follows from the equation that $dM_x/dt \neq 0$ that is all three components of the magnetization exist. We need to introduce the following spherical components of the magnetization

$$M_z = M \cdot \sin\theta \cdot \cos\varphi,$$
$$M_y = M \cdot \sin\theta \cdot \sin\varphi, \qquad , \qquad (A7)$$
$$M_x = M \cdot \cos\theta,$$

where $\theta$ is the angle between vectors $\mathbf{M}$ and $\mathbf{x}$. In the initial stationary state the angle $\theta = \theta_0 = \pi/2$. We seek the solution in the form:
$\theta = \theta_0 + \theta_1 \exp(pt)$ and $\varphi = \varphi_0 + \varphi_1 \exp(pt)$, where $\theta_1, \varphi_1 \ll 1$. Then we get after linearization

$$\left(p + \frac{\alpha_d \gamma}{M} \cdot \frac{\partial^2 U_{tot}}{\partial \theta_0^2}\right)\theta_1 + \frac{\gamma}{M} \cdot \frac{\partial^2 U_{tot}}{\partial \varphi_0^2}\varphi_1 = 0$$
$$-\frac{\gamma}{M} \cdot \frac{\partial^2 U_{tot}}{\partial \theta_0^2}\theta_1 + \left(p + \frac{\alpha_d \gamma}{M} \cdot \frac{\partial^2 U_{tot}}{\partial \varphi_0^2}\right)\varphi_1 = 0 \qquad (A8)$$

To derive the equation (A8) we have used the following two conditions of the stationary state: 1) the condition (A6) and 2) the condition $\partial U_{tot}/\partial \theta_0 = 0$, the latter condition being got directly from the definition of our demagnetization energy taken at the point $\theta_0 = \pi/2$.

We get from (A8) the following characteristic roots:

$$p_{1,2} = -\frac{\alpha_d \gamma}{M} \cdot \left(4\pi M^2 + \frac{\partial^2 U}{\partial \varphi_0^2}\right) \pm i\gamma \sqrt{4\pi \frac{\partial^2 U}{\partial \varphi_0^2}}. \qquad (A9)$$

These roots show:
1) the stability is ensured for

$$\frac{\partial^2 U}{\partial \varphi_0^2} > 0; \qquad (A10)$$

2) the fluctuations damp off due to dissipation, decrement being proportional to $\alpha_d$, if the condition (A10) is satisfied;

3) the fluctuations rise rapidly if the condition (A10) is not satisfied even in the absence of the dissipation.

The derived relations (A6) and (A10) represent, of course, the energy minimum conditions. These conditions justify our approach, which uses energy minimization principle in spite of the current. Why may it be possible? Apparently the magnetic subsystem with the total energy $U_{tot}$ is in a partially equilibrium state because the current do not produce any additional irreversible processes in the subsystem.